\documentclass[10pt,conference]{IEEEtran}
\usepackage{xcolor}
\usepackage{booktabs}
\usepackage{fontawesome}
\usepackage{subfigure}
\usepackage{bm}
\usepackage{xspace}
\usepackage{float}
\usepackage{graphicx}
\usepackage{accents}
\usepackage{threeparttable}
\usepackage{diagbox}
\usepackage{textcomp}
\usepackage[skins]{tcolorbox}
\usepackage{enumitem}
\usepackage{graphicx}
\usepackage{multirow}
\usepackage{longtable}
\usepackage{pgf-pie} 
\usepackage{cite}
\usepackage{amssymb,amsfonts}
\usepackage{algorithmic}
\usepackage{url}
\usepackage[hidelinks]{hyperref}

\def\BibTeX{{\rm B\kern-.05em{\sc i\kern-.025em b}\kern-.08em
    T\kern-.1667em\lower.7ex\hbox{E}\kern-.125emX}}
    
\newcommand{\myfancylabel}{\begin{tikzpicture}[every node/.style={rotate=45}]
\node[fill,inner sep=0pt,minimum size=0.5ex] at (0ex,0.5ex) {};
\node[fill,inner sep=0pt,minimum size=0.5ex] at (0ex,-0.5ex) {};
\node[fill,inner sep=0pt,minimum size=0.5ex] at (0.5ex,0ex) {};
\node[fill,inner sep=0pt,minimum size=0.5ex] at (-0.5ex,0ex) {};
\end{tikzpicture}}
\newcommand{\Category}[1]{{\sffamily #1}\xspace}

\newlength{\fsize}

\tcbset{
  my box/.style={
    enhanced,
    colframe=#1!80,
    colback=#1!10,
    attach boxed title to top left={xshift=0.2cm, yshift=-0.2cm},
    boxed title style={
      colback=#1!80,
      outer arc=0pt,
      arc=0pt,
      top=0pt,
      bottom=0pt,
    },
  },
}
\newtcolorbox{result-rq}[1]{
  my box=black,
  title=#1,
  boxrule=1.2pt,top=6pt,bottom=3.5pt,left=6pt,right=6pt
}

\begin{document}
\title{\vspace{-1.2cm} \textcolor{gray}{\footnotesize This is the author's version of the paper that has been accepted for publication in\\ \vspace{-0.7cm} the 38th IEEE/ACM International Conference on Automated Software Engineering (ASE 2023)}\\ \vspace{0.3cm}
Personalized First Issue Recommender for Newcomers in Open Source Projects}

\author{
    \IEEEauthorblockN{Wenxin Xiao\IEEEauthorrefmark{1}, Jingyue Li\IEEEauthorrefmark{2}, Hao He\IEEEauthorrefmark{1}, Ruiqiao Qiu\IEEEauthorrefmark{1}, Minghui Zhou\IEEEauthorrefmark{1}\IEEEauthorrefmark{3}}
    \IEEEauthorblockA{\IEEEauthorrefmark{1}School of Computer Science, Peking University, China\\
    Key Laboratory of High Confidence Software Technologies, Ministry of Education, China\\
    wenxin.xiao@stu.pku.edu.cn, heh@pku.edu.cn, qrq2001@stu.pku.edu.cn, zhmh@pku.edu.cn}
    \IEEEauthorblockA{\IEEEauthorrefmark{2}Norwegian University of Science and Technology, Trondheim, Norway
    \\jingyue.li@ntnu.no}
}

\maketitle

\begingroup\renewcommand\thefootnote{\IEEEauthorrefmark{3}}
\footnotetext{Minghui Zhou is the corresponding author.}
\endgroup

\thispagestyle{plain}\pagestyle{plain}
\begin{abstract}
Many open source projects provide \textit{good first issues} (GFIs) to attract and retain newcomers.
Although several automated GFI recommenders have been proposed, existing recommenders are limited to recommending generic GFIs without considering differences between individual newcomers. 
However, we observe mismatches between generic GFIs and the diverse background of newcomers, resulting in failed attempts, discouraged onboarding, and delayed issue resolution.
To address this problem, we assume that personalized first issues (PFIs) for newcomers could help reduce the mismatches. 
To justify the assumption, we empirically analyze 37 newcomers and their first issues resolved across multiple projects.
We find that the first issues resolved by the same newcomer share similarities in task type, programming language, and project domain. 
These findings underscore the need for a PFI recommender to improve over state-of-the-art approaches. 
For that purpose, we identify features that influence newcomers' personalized selection of first issues by analyzing the relationship between possible features of the newcomers and the characteristics of the newcomers' chosen first issues. 
We find that the expertise preference, OSS experience, activeness, and sentiment of newcomers drive their personalized choice of the first issues.
Based on these findings, we propose a Personalized First Issue Recommender (PFIRec), which employs LamdaMART to rank candidate issues for a given newcomer by leveraging the identified influential features. 
We evaluate PFIRec using a dataset of 68,858 issues from 100 GitHub projects. 
The evaluation results show that PFIRec outperforms existing first issue recommenders, potentially doubling the probability that the top recommended issue is suitable for a specific newcomer and reducing one-third of a newcomer's unsuccessful attempts to identify suitable first issues, in the median.
We provide a replication package at \url{https://zenodo.org/record/7915841}.
\end{abstract}

\section{Introduction}
\label{sec:intro}

Open source software (OSS) project developers face challenges when it comes to selecting suitable issues to contribute to, particularly when they are new to a project~\cite{shibuya2009understanding}.
If a newcomer's initial attempts to resolve their chosen issues fail, it can waste their and the maintainers' time and effort, delaying issue resolution and blocking opportunities for other newcomers.
To address this problem, GitHub encourages OSS projects to label \textit{good first issues} (GFIs) to highlight suitable tasks for project newcomers.

To reduce projects' effort to label GFIs, several automated GFI recommending approaches and tools are proposed and have achieved reliable performance~\cite{DBLP:conf/icsm/StanikMMFM18,DBLP:conf/esem/HuangWWLWW21,DBLP:conf/icse/XiaoHXTDZ22,DBLP:conf/sigsoft/HeSXHZ22}. 
However, they treat newcomers as an undifferentiated mass without considering the individual differences among contributors, and such generic GFIs may not match newcomers' diverse backgrounds, leading to unsuccessful resolution attempts.
For example, in our dataset (see Section~\ref{sec:dataset} for more details), we identified 199 GFIs wherein 238 newcomers actively engaged by providing comments on the issues or participating in issue events~\cite{Issueeve24:online}. These interested newcomers were potential GFI resolvers. 
However, 224 (94.1\%) of the newcomers did not resolve the GFIs they were involved in, and these GFIs were resolved by other newcomers who did not engage in commenting or participating in issue events. 
Merely 14 (5.9\%) of the 238 newcomers resolved the GFIs they participated in. 

Furthermore, we encountered instances that vividly portray the divergence between specific newcomers and GFIs:
\begin{itemize}
    \item In Issue \#151196 of VS Code~\cite{Supportb42:online}, a PR made by a newcomer was closed 26 days after he claimed the issue due to his lack of experience with the TextMate grammar necessary to resolve the issue; only four days later, the issue was successfully resolved by another newcomer. 
    \item In Issue \#50207 of the pandas project~\cite{CIchecke75:online}, a newcomer stated, 36 days after claiming to take the issue, that he did not have enough time to resolve the issue; then, a second newcomer took five days to resolve the issue. 
\end{itemize}
The two examples show that GFIs are not generally suitable for arbitrary newcomers and that newcomers may be unable to resolve some GFIs if they lack the necessary knowledge or skills or sufficient time. 
Some projects have specific requirements for newcomers' experience, which makes it even more necessary to match the background of newcomers to the requirements of the projects and GFIs. For instance, the Bitcoin Core project states that \textit{a good first issue isn't targeted toward software development beginners. At the very least, you'll need basic Git proficiency and ideally C++ and/or Python proficiency, too, given that the Bitcoin Core codebase is written in these languages}~\cite{OpEdWant22:online}.

Therefore, we assume that personalized first issue (PFI) recommendations, based on the historical behavior of each newcomer, may significantly improve the effectiveness of GFI recommendation tools and therefore, reduce the failed attempts of project newcomers. 
To justify the assumption, we posit the first research question (RQ) to check whether each newcomer exhibits distinct first-issue preferences across projects and provide a theoretical underpinning for the implementation of PFI recommenders:
\begin{itemize}
    \item \textbf{RQ1:} Does each OSS newcomer exhibit personalized preferences when choosing the first issue in a new project?
\end{itemize}
To answer RQ1, we identify 37 newcomers contributing to more than two projects from the GFI-Bot dataset~\cite{DBLP:conf/sigsoft/HeSXHZ22} and analyze the 123 first issues\footnote{We will abbreviate the first issue that newcomers resolve in a project as FI hereafter.} they resolved.
We then compare the FI characteristics (i.e., task types, programming languages, and project domains) for each newcomer across projects as well as across newcomers. 
The results show that the FI characteristics are similar across projects for each individual but different among newcomers. 
This indicates that newcomers' choice of the FIs in new projects is mostly personalized, and it will be beneficial to propose PFIs to project newcomers.   

To propose a PFI recommender, we need to know which features of newcomers influence their personalized choice of FIs. Thus, we pose the second research question:
\begin{itemize}
    \item \textbf{RQ2:} Which features of newcomers are related to their choices of the first issues in projects?
\end{itemize}
To answer RQ2, we analyze the relationship between possible influential features and the characteristics of the FIs selected by the 37 newcomers. We find that their resolved FIs are more semantically akin to their historically resolved ones than those resolved by other newcomers, and the newcomers' experience, activeness, and sentiment features impact their FI choices. 

Based on the identified features from RQ2, we propose PFIRec, a personalized issue recommendation approach that ranks issues for specific newcomers using the LambdaMART model~\cite{burges2010ranknet}. To train the model and evaluate the effectiveness of PFIRec, we construct a dataset by collecting resolver (i.e., newcomer) features and issue features of 11,615 FIs and their 57,243 contemporary issues from 100 GitHub projects aiming to attract newcomers. 
The experimental results show that PFIRec outperforms state-of-the-art GFI recommenders. Compared with the best state-of-the-art GFI recommender, PFIRec can double the probability that the top recommended issue is suitable for a specific newcomer and reduce one-third of the number of a newcomer's unsuccessful attempts to find appropriate issues, in the median.

This paper makes the following contributions:
\begin{itemize}
\item We give empirical evidence that newcomers' choices of the first issues in new projects are mostly personalized. To the best of our knowledge, this is the first study to explore personalized recommendations for the first issue.
\item We provide insights that historical contributions, general OSS experience, activeness, and sentiment of newcomers can influence their personalized choice of first issues.
\item We propose and evaluate PFIRec, a personalized issue recommender that ranks issues for specific newcomers, outperforming state-of-the-art GFI recommenders.
\end{itemize}
 
The rest of the paper is organized as follows.
Section~\ref{backgroundandmoti} 
introduces background and related work. 
The methodology and results of the empirical study (answering RQ1 and RQ2) are presented in Section~\ref{empirical}. 
PFIRec's design, implementation, and evaluation are in Section~\ref{rq3}.
Section~\ref{discussion} discusses comparison with related work and threats to validity.
Section~\ref{conclusion} concludes the paper. 

\section{Background and Related Work}\label{backgroundandmoti}

\subsection{Newcomer Onboarding}

OSS projects depend on many volunteers and their contributions to evolve and meet various requirements~\cite{mockus2002two}.
However, the sustainability of OSS projects is at risk due to the loose organizational structure, which can lead to the potential loss of contributors~\cite{2012ICSE-Zhou-What}. 
For example, the disclosure of HeartBleed and Log4Shell vulnerabilities brought attention to the fact that the maintenance of OpenSSL and Log4j relied on contributions from only two and four developers, respectively~\cite{6VolkanY51:online,HeartBleed-Two-Guy}, causing considerable damage to numerous dependent projects~\cite{GoogleOn56:online,Heartble43:online}. 
To ensure the sustainability of OSS projects and overcome the unmaintained status, consistently attracting and retaining newcomers is crucial~\cite{2020FSE-Tan-First,DBLP:conf/sigsoft/CoelhoV17}. 

Despite motivations, such as learning software development skills and pursuing future career prospects, which motivate newcomers to participate in OSS projects~\cite{10.5555/776816.776867,Krogh2012Carrots}, they often face barriers, such as lack of project support and the necessary skills\cite{steinmacher2014barriers,shibuya2009understanding}, making successful onboarding challenging. Multiple failed attempts may cause newcomers to abandon their contributions, which wastes not only their time and effort but also those of project maintainers.

Numerous studies explore the barriers newcomers face when contributing to new projects, along with strategies to overcome these challenges. Steinmacher et al.~\cite{2019Softw-Igor-Let} provide onboarding guidelines, such as creating a newcomer portal. 
Jensen et al.~\cite{2011HICSS-Carlos-Joining} emphasize the importance of timely responses to newcomers. 
Santos et al.~\cite{DBLP:conf/esem/SantosTPWSSG22} indicate that newcomers want more onboarding support, Tan et al.~\cite{tan2023enough} find expert involvement helpful, and He et al.~\cite{DBLP:conf/icse/HeZWL23} show that courses can be an effective intervention. Zhou and Mockus~\cite{DBLP:journals/tse/ZhouM15} recommend mentoring potential long-term contributors, and Panichella~\cite{DBLP:conf/icsm/Panichella15} proposes a tool to match suitable mentors with newcomers.

Despite the above strategies, one critical challenge remains, i.e., identifying suitable tasks for newcomers who may lack the requisite skills and knowledge~\cite{DBLP:conf/hicss/SteinmacherCG15,DBLP:conf/vissoft/ParkJ09}.
Wang and Sarma~\cite{DBLP:conf/icse/WangS11} present an interactive tool to visualize bug descriptions for newcomers.
However, it is still difficult for newcomers to find suitable tasks among a massive number of tasks.
To address these challenges, \textit{good first issue} labels are applied to highlight opportunities for newcomers to contribute~\cite{GitHub-GFI}. 

\subsection{Newcomer Task Recommendation}\label{relatedwork}
Several studies investigate the current status and limitations of the GFI mechanism. Horiguchi et al.~\cite{DBLP:conf/wcre/HoriguchiOO21} show that GFIs are resolved by less experienced developers compared to regular issues, while Alderliesten and Zaidman~\cite{DBLP:conf/icse/AlderliestenZ21} reveal that only a small percentage of GFIs are addressed by newcomers. Xiao et al.~\cite{xiao2023early} discover that the GFI mechanism has limited effectiveness in the early stages of OSS projects.
Tan et al.~\cite{2020FSE-Tan-First} find that manually labeled GFIs are often highly insufficient, and cognitive mismatch between newcomers and veterans can lead to inappropriate GFIs labeled by veterans. To overcome the limitations of the GFI mechanism due to sparse and subjective labeling, automated approaches are proposed to recommend issues to newcomers~\cite{DBLP:conf/esem/HuangWWLWW21,DBLP:conf/icse/XiaoHXTDZ22,DBLP:conf/sigsoft/HeSXHZ22,DBLP:conf/icsm/StanikMMFM18}. Among them, the most efficient automated approach is GFI-Bot~\cite{DBLP:conf/sigsoft/HeSXHZ22}, which is developed as a GitHub bot. These automated approaches extract various features to identify issues that may be suitable for any newcomer. 
However, these approaches do not take into account the diverse backgrounds of newcomers. When using these tools, newcomers may be recommended issues that do not match their preferences or that they find difficult or uninteresting to resolve.

Personalized recommendation have been explored in other software engineering tasks. For example, Anvik et al.~\cite{DBLP:journals/tosem/AnvikM11}, Ashok et al.~\cite{DBLP:conf/sigsoft/AshokJLRSV09}, and MacDonald et al.~\cite{DBLP:conf/cikm/MacdonaldO06} recommend experts to a specific task based on their familiarity with task-related artifacts or source code.
Wang et al.~\cite{DBLP:conf/icse/WangY0HWW20} recommend crowdworkers with relevant experience to test tasks on crowdsourcing platforms using text term matching.
Ye et al.~\cite{DBLP:conf/kbse/Ye0WW18} and Zhou et al.~\cite{DBLP:conf/icws/ZhouWS21} propose approaches to personally recommend collaborative teammates and GitHub repositories for developers based on their skill proficiency and collaboration history, and topic similarity. 
In general, previous research on matching developers and tasks focuses on recommending experts who are familiar with the task rather than newcomers who do not have experience in the project. Other personalized recommendation studies recommend collaborators or projects instead of tasks to developers. These approaches cannot be used directly to resolve our problem.

\section{Empirical study}\label{empirical}
As previously stated, generic GFIs may not align with newcomers' individual abilities and interests, and personalized recommendations have been applied in software engineering to improve the efficacy of recommendations. Therefore, we are motivated to answer RQ1 and RQ2 to empirically investigate whether it is rational to recommend personalized FIs and what factors influence newcomers' personalized choices. 

\subsection{RQ1 Methodology}\label{methrq1}
According to previous studies (e.g., ~\cite{Whatmake97:online,observations:online,2020FSE-Tan-First,DBLP:conf/icse/XiaoHXTDZ22}), certain types of tasks, such as document changes, are suitable for newcomers. Additionally, project background, including programming language and project domain, may serve as an indicator of whether newcomers can resolve the issues~\cite{DBLP:conf/icse/XiaoHXTDZ22,DBLP:conf/sigsoft/HeSXHZ22}. Hence, we aim to compare the issues resolved by newcomers based on task type, programming language, and project domain to investigate whether there are variations among issues resolved by different newcomers and whether the FIs resolved by the same newcomer in different projects are similar.

\subsubsection{Data Collection}\label{datacollection-RQ1}
To answer RQ1, we need a dataset containing issue- and resolver-related data collected from projects that are attractive to and have already onboarded many newcomers.
The GFI-Bot dataset~\cite{DBLP:conf/sigsoft/HeSXHZ22}, which is currently the largest and most up-to-date dataset for recommending issues to newcomers, has collected 159,919 issues until June 2022 from the 100 GitHub projects with the highest number of \textit{good first issues}, including prestigious projects, such as pandas and VS Code~\cite{DBLP:conf/icse/XiaoHXTDZ22}. These projects are willing to attract newcomers and are highly desirable for newcomers to join.
The dataset contains issues, reporter and project owner data. We identified 12,755 newcomers and the 13,266 FIs they resolved from the dataset to conduct this study. 

Investigating FIs resolved by the same newcomer across multiple projects can help us understand the newcomers' tendency to select issues in new projects. 
However, insufficient records of a newcomer's FIs make it challenging to explore their inclination toward new project issues. 
Among the 12,755 newcomers, 12,293 newcomers addressed FIs in one project, while 425, 26, 10, and one newcomer resolved FIs in two, three, four, and five projects (considering only the 100 projects in the dataset), respectively. 
We manually label and analyze the 123 (=26$\times$3+10$\times$4+1$\times$5) FIs resolved by the 37 newcomers who participated in more than two projects to answer RQ1.

\subsubsection{Data Analysis}\label{dataanalysis-RQ1}
We first manually categorize the 123 FIs based on \textbf{task type}, which encompasses two kinds of categories: 1) code/document tasks; and 2) corrective/adaptive/perfective tasks, following the widely adopted software change classification proposed by Mockus and Votta~\cite{DBLP:conf/icsm/MockusV00}.
We then record the primary \textbf{programming languages} of the projects, which involve JavaScript, Python, Java, TypeScript, C\#, PHP, and C++. 
Finally, we label the \textbf{project domains} of issues with six categories, consistent with the classification of GitHub project domains proposed by Borges et al.~\cite{DBLP:conf/icsm/BorgesHV16}: \textit{application software}, \textit{system software}, \textit{web libraries and frameworks}, \textit{non-web libraries and frameworks}, \textit{software tools}, and \textit{documentation}. 
The classification of GitHub repositories is widely recognized and used~\cite{DBLP:journals/corr/abs-2208-00269,DBLP:journals/jss/BorgesV18}.
Two authors independently mark these issues. The inter-rater reliability between the two authors is assessed through Cohen's Kappa ($\kappa$), which yielded values of 0.98, 0.86, 1, and 0.91 for task type (code/document), task type (corrective/adaptive/perfective), programming language, and project domain, respectively. To resolve any conflicts that arise during the labeling process, an arbitrator with top-tier software engineering conference paper publications and four years of OSS development experience is involved until all labels are agreed upon within the labeling group. We analyze the proportion of different issue types to investigate the tendency of each of the 37 newcomers toward the FIs.

\subsection{RQ2 Methodology}\label{rq2method}

The reinforcement theory~\cite{ferster1957schedules} states, \textit{behavior is driven by its consequences and positive reinforcements (rewards) serve as facilitators that increase the likelihood of the desired behavior's recurrence}. 
Based on this theory, we suspect that newcomers tend to select FIs that are similar to their historical contributions to replicate their successes.
Moreover, newcomers with different OSS experiences may have varying preferences in terms of issue types and difficulty levels.
Thus, we investigated features that may influence newcomers' choices of FIs. Table~\ref{tab:reasons} shows the feature groups and the rationale for choosing them. 

\begin{table}[h!t]
\scriptsize
\centering
\vspace{-0.3cm}
\caption{The feature groups we choose and the justifications for choosing them}
\vspace{-0.2cm}
\label{tab:reasons}
\begin{tabular}{l l}
\toprule
Feature Groups&\multirow{2}{5.3cm}{Why can the feature group possibly influence the newcomers' choice of FIs?}\\
\\
\midrule
\multicolumn{2}{l}{\textbf{Expertise Preference}}\\
Content Preference  &\multirow{3}{5.3cm}{Features of this dimension can reflect similarities between newcomers' historical contributions and candidate issues regarding their task types and programming languages.}\\
\\
\\
\\
Domain Preference &\multirow{2}{5.3cm}{Features of this dimension can gauge newcomers' preference for the relevant project domain.}\\
\\
\midrule
\textbf{General OSS Experience}&\multirow{2}{5.3cm}{Newcomers with more OSS experience may exhibit a preference for tackling more challenging tasks.}\\
\\
\midrule
\textbf{Activeness}&\multirow{6}{5.3cm}{The resolution of specific types of issues and tasks may require a prolonged period of time. Activeness measures how active the newcomers have been on GitHub recently. According to Wang et al.~\cite{DBLP:conf/icse/WangY0HWW20}, activeness, i.e., recent activity, can be used to measure the availability of newcomers.}\\
\\
\\
\\
\\
\\
\midrule
\textbf{Sentiment}&\multirow{7}{5.3cm}{Studies~\cite{DBLP:conf/icse/XiaoHXTDZ22, tan2023enough} reveal that interactions with experienced developers can be advantageous for newcomers in resolving issues. We posit that newcomers who demonstrate a more positive communication style are more likely to receive aid from experienced developers when grappling with complex issues.}\\
\\
\\
\\
\\
\\
\\
\bottomrule
\end{tabular}
\vspace{-0.2cm}
\end{table}

\subsubsection{Data Collection}\label{datacollection-RQ2}
We use the same labeled data from the 123 FIs resolved by the 37 newcomers in our analysis of RQ1. We collect the following features of newcomers regarding their expertise preference, general OSS experience, activeness, and sentiment. 

\textbf{Expertise preference features.}
As explained in Table~\ref{tab:reasons}, we use \textit{content preference} and \textit{domain preference} to measure the FI preference of newcomers in terms of task type, programming language, and project domain.

\textit{\textbf{Textual data collection and preprocessing to calculate content and domain preference.}}
We calculate \textit{content preference} by comparing the title and description of candidate issues with reported issues, resolved issues, PRs, and commit messages made by newcomers in history. 
Similarly, we calculate \textit{domain preference} by comparing READMEs, topics, descriptions of candidate issues' projects and the projects that the newcomer has contributed to.

For issues and PRs, we first concatenate their title and description and then remove URLs, code snippets, and images to extract plain text. Next, we reduce noise by applying textual document processing steps, including word segmentation, lemmatization, and stop word filtering, which are commonly used in software engineering studies~\cite{DBLP:conf/icse/WangY0HWW20,DBLP:journals/access/HuangLWRCZ19}. Thus, we can calculate Jaccard similarity between the texts.

We also calculate cosine similarity between texts using pre-trained language models like bidirectional encoder representations from transformers (BERT)~\cite{DBLP:conf/naacl/DevlinCLT19}. However, the embeddings created by BERT are highly anisotropic and semantically non-smoothing, making it difficult to apply directly to similarity metrics~\cite{DBLP:conf/emnlp/GaoYC21,DBLP:conf/emnlp/LiZHWYL20}. Simple contrastive learning of sentence embeddings (SimCSE)~\cite{DBLP:conf/emnlp/GaoYC21} is proposed to address this problem, which regularizes the pre-trained embeddings in an anisotropic space by optimizing a contrastive learning objective. This encourages similar embeddings to be closer in the embedding space and dissimilar embeddings to be farther apart, making the embeddings more uniform. We adopt SimCSE to extract features from texts using the uncased SimCSE model~\cite{princeto91:online} to tokenize and pad the texts and generate embeddings of the texts with pooler output.

\textit{\textbf{Calculating content preference.}} 
Following textual data preprocessing, we utilize Jaccard and cosine similarity to measure the similarity of the used terms and the semantic similarity of texts, respectively. For example, the Jaccard and cosine similarities between a newcomer's cumulative PRs and the candidate issue are calculated as follows. 
        
\myfancylabel ~\textit{Cumulative PR cosine similarity} calculates and accumulates cosine similarity between the embedding vectors of each historical PR made by newcomer $d$ and candidate issue $s$:
\begin{equation}
\footnotesize
PRCos(d,s) = \sum\limits_{r=1}^{R}\frac{\sum\limits_{n=1}^{N}PRemb_{r,n}(d)ISSemb_n(s)}{\sqrt{\sum\limits_{n=1}^{N}PRemb^2_{r,n}(d)}\sqrt{\sum\limits_{n=1}^{N}ISSemb^2_n(s)}}
\label{eqcos}
\end{equation}
where $R$ is the number of PRs, $N$ is number of dimensions of embedding vector, $PRemb_{r,n}(d)$ is the $n$\textsuperscript{th} dimensional value of embedding vector of the $r$\textsuperscript{th} PR made by newcomer $d$, and $ISSemb_n(s)$ is the $n$\textsuperscript{th} dimensional value of embedding vector of candidate issue $s$.

\myfancylabel ~\textit{Cumulative PR Jaccard similarity} calculates the sum of Jaccard similarity between historical PRs from newcomer $d$ and candidate issue $s$:
\begin{equation}
\footnotesize
PRJac(d,s) = \sum\limits_{r=1}^{R}\frac{A_r(d)\cap B(s)}{A_r(d)\cup B(s)},\label{eqjac}
\end{equation}
where $A_r(d)$ is the term set of preprocessed text of the $r$\textsuperscript{th} PR made by newcomer $d$, $B(s)$ is the term set of preprocessed text of issue $s$, $\cap$ represents intersection and $\cup$ denotes union.

Besides calculating the cosine and Jaccard similarities, we also calculate the cumulative number of relevant PRs. 

\myfancylabel ~\textit{Cumulative number of relevant PRs} count the number of PRs from newcomer $d$ that have common labels with the candidate issue $s$:
\begin{equation}
\footnotesize
PRLab(d,s) = \sum\limits_{r=1}^{R}int((PRlabels_r(d)\cap Isslabels(s))\neq\emptyset),
\end{equation}
where $PRlabels_r(d)$ denotes the label set of the $r$\textsuperscript{th} PR made by newcomer $d$, and $Isslabels(s)$ is the label set of candidate issue $s$.

Then we calculate the average values of these measurements by dividing the cumulative values by the number of PRs: $PRJac/Cos/Lab\_ave(d,s)=\frac{PRJac/Cos/Lab(d,s)}{R}$. 
The cumulative measurements imply whether the newcomer has acquired sufficient experience in the relevant content, while the average similarity measurements showcase the overall similarity between the newcomer's historical contributions and the candidate issue.

We apply the same similarity measurements as for PRs to analyze the similarity of historical commit messages, resolved issues, reported issues, commented issues, and reviewed PRs to candidate issues, except for commit messages lacking label-related data.

We assess the newcomer's familiarity with the programming language of the candidate issues by counting the number of commits they have made in prior projects that share the same primary programming language as the candidate issue. 

\textit{\textbf{Calculating domain preference.}}
We collect the descriptions and READMEs of the projects corresponding to the candidate issues and the historical commits of the newcomers. Then we calculate the aforementioned similarity measurements between the descriptions and the READMEs, except for the label-related measures. 

To measure the newcomers' accumulated experience on the topics of the candidate issues, we record the number and ratio of commits made by them in projects that share at least one common topic with the project to which the candidate issue belongs.

\textbf{General OSS experience features.}
According to Argote and Miron-Spektor ~\cite{DBLP:journals/orgsci/ArgoteM11}, developer experience can be measured by the cumulative number of tasks performed. 
We measure the general cumulative OSS experience of newcomers through the quantification of their contributions, including commits, PRs, PR reviews, repositories contributed to, and issues reported. 
Additionally, we consider the number of issues they have reported in the new project as an indication of their interest.

\textbf{Activeness features.}
To gauge newcomer activeness, we count the number of their recent contributions within the past one, two, and three months, including the number of commits, PRs, and reported issues on GitHub.

\textbf{Sentiment features.}
We calculate sentiment measures for newcomers with the Python TextBlob package~\cite{textblob} using historical PRs' and issues' titles and descriptions. The measurements include the mean and median textual sentiment polarity values of their historically contributed PRs and reported issues.

\subsubsection{Data Analysis}\label{dataanalysis-RQ2} We first make pairwise comparisons between each FI and the issues resolved by the same newcomer before resolving the FI, as well as between each FI and the FIs resolved by the other 36 newcomers. The comparisons are conducted by calculating the cosine similarity and Jaccard similarity of the issues' texts and corresponding project descriptions. By doing so, we measure both the content and domain similarity of the issues to explore whether the newcomers tend to choose FIs that are similar to their historical contributions.
We then collect features of the general OSS experience, activeness, and sentiment groups for the 37 newcomers.
We compare the distribution of the features corresponding to FIs of different task types and domains to explore the relationship between these features and the FIs.

\subsection{RQ1 Results: Newcomers' Different FI Preferences}
\begin{figure}
\centering
\includegraphics[scale=0.42,trim=0 0 0 0]{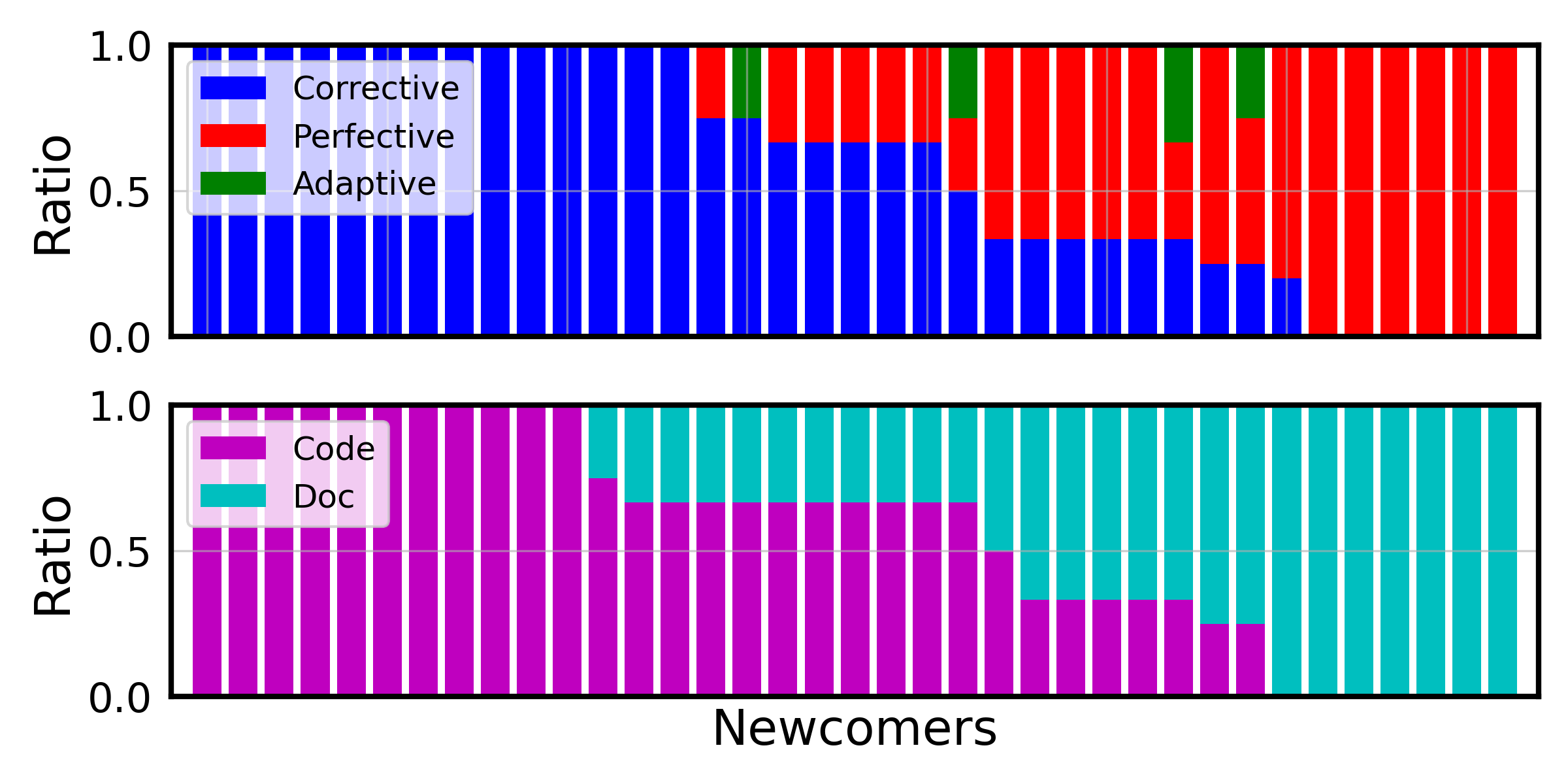}
\vspace{-0.3cm}
\caption{Task type distribution of FIs}
\vspace{-0.3cm}
\label{Task type}
\end{figure}

\begin{figure}
\centering
\includegraphics[scale=0.42,trim=0 0 0 0]{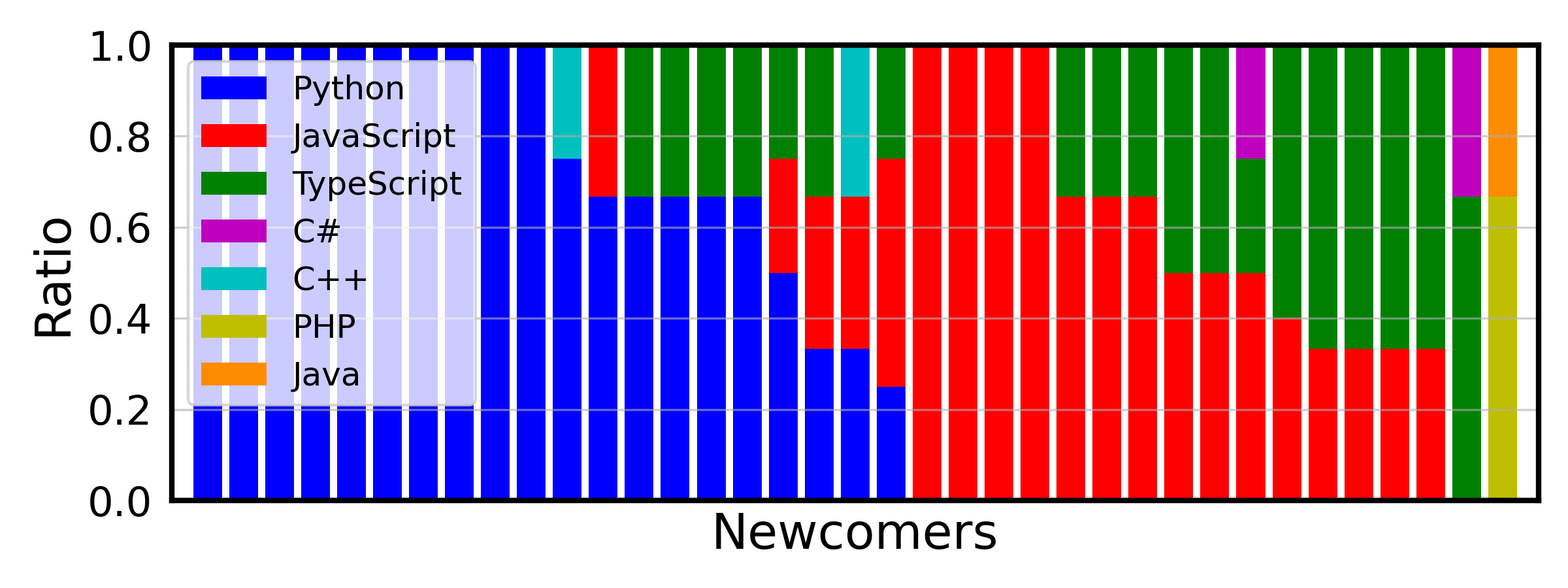}
\vspace{-0.3cm}
\caption{Programming language distribution of FIs}
\vspace{-0.3cm}
\label{language}
\end{figure}
\begin{figure}
\centering
\includegraphics[scale=0.42,trim=0 0 0 0]{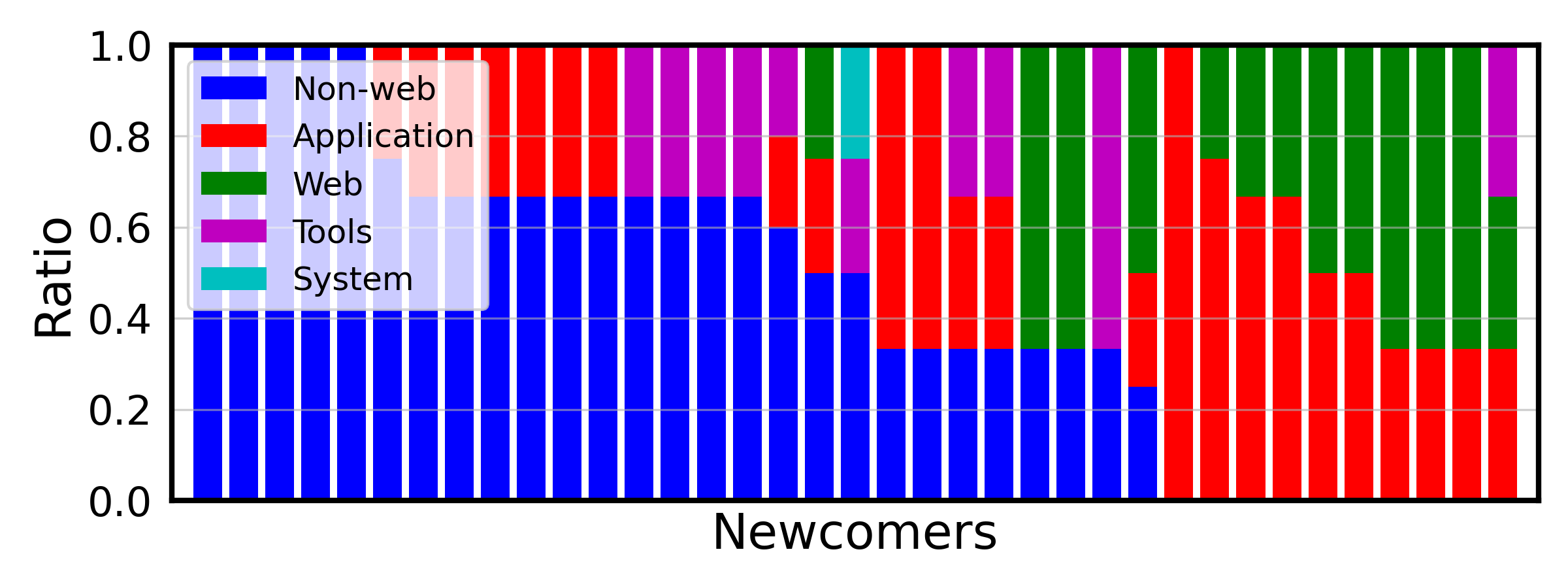}
\vspace{-0.25cm}
\caption{Project domain distribution of FIs}
\vspace{-0.3cm}
\label{domain}
\end{figure}

Figure~\ref{Task type} to~\ref{domain} show the percentage of different types of FIs resolved by each newcomer, which offer insights into their FI preferences. Each of the 37 bars depicted in each figure corresponds to one of the newcomers.

\textbf{Task type}. Manual labeling of the 123 FIs shows that 69 FIs are code related, whereas only 54 FIs are document related. Figure~\ref{Task type} reveals the task type distribution of the issues resolved by the 37 newcomers, demonstrating that 11 newcomers exclusively resolved code-related issues, while seven newcomers dedicated their efforts solely to document-related issues in new projects. This finding suggests that simply recommending document-related issues to newcomers may not be an optimal strategy. Figure~\ref{Task type} also shows that 20 newcomers focused on a single type of corrective, adaptive, or perfective issue, indicating their consistent preference for task type in new projects. 

\textbf{Programming language}. As depicted in Figure~\ref{language}, among the 37 newcomers, ten exclusively addressed Python FIs, four exclusively resolved JavaScript FIs, and ten only resolved JavaScript and TypeScript FIs. Notably, no newcomer submitted initial contributions in more than three programming languages, indicating a preference for leveraging their expertise in specific programming languages when contributing.

\textbf{Project domain}. Figure~\ref{domain} shows that most newcomers concentrated their efforts on particular domains, with six newcomers focusing on addressing FIs from projects in a single domain and 24 newcomers dedicating their efforts to resolving issues within two domains. Only seven newcomers resolved FIs across three domains, and none distributed their efforts across more than three domains. Note that none of the 123 FIs belong to documentation projects.

\begin{result-rq}{Summary for RQ1:}
Regarding task type, programming language, and project domain, newcomers tend to exhibit consistent preferences in their first issues of new projects.
\end{result-rq}

\subsection{RQ2 Results: Newcomer Features Influencing FI choices}
\begin{figure}
\centering
\subfigure[Distribution of similarity measurements (issues' cosine/Jaccard similarity, project descriptions' cosine/Jaccard similarity)
]
{
\centering
\includegraphics[scale=0.42]{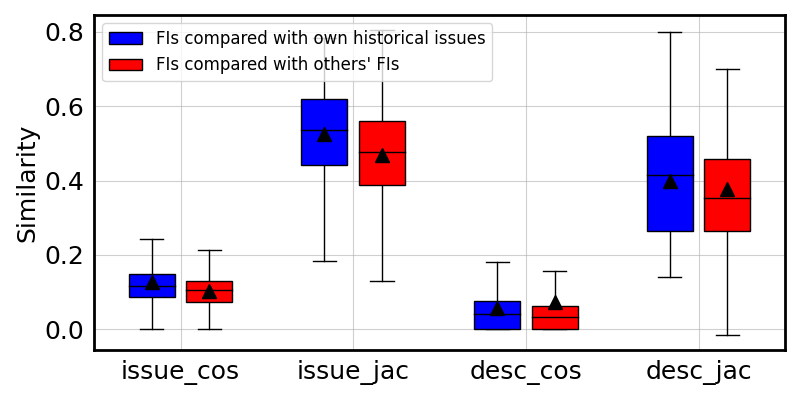}
\label{experience}
}
\subfigure[Distribution of newcomer features across different task types]
{
\centering
\includegraphics[scale=0.42]{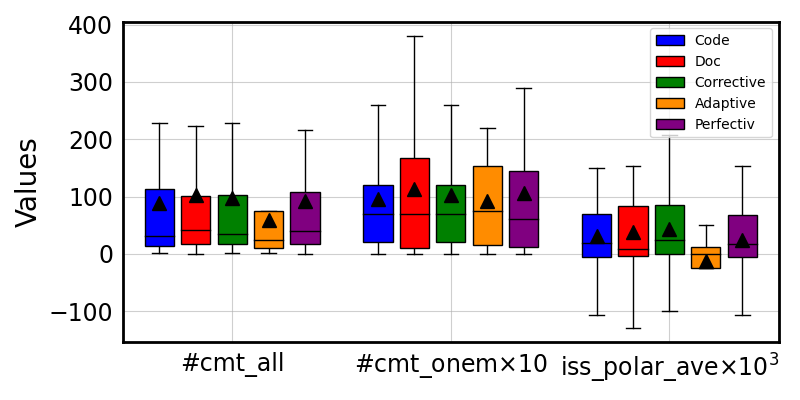}
\label{typedis}
}
\subfigure[Distribution of newcomer features across different project domains]
{
\centering
\includegraphics[scale=0.42]{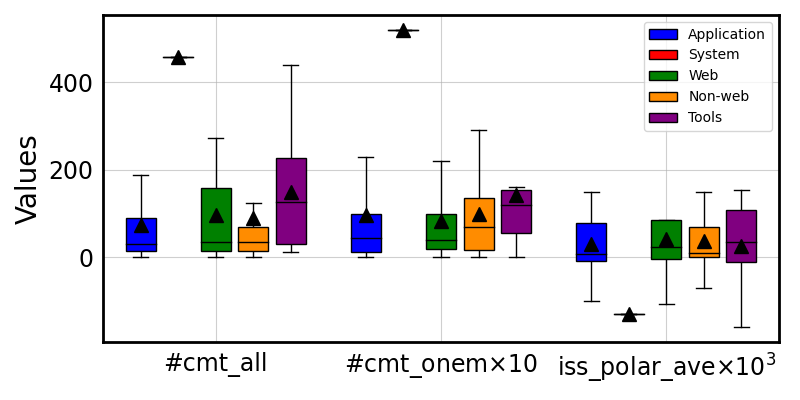}
\label{domaindis}
}
\caption{Example of features that are related to newcomer FI preferences.}
\vspace{-0.3mm}
\label{figevaluation}
\end{figure}

\textbf{Newcomers' content and domain preference influence their FI selection.} 
Figure~\ref{experience} demonstrates that the FIs chosen by the newcomers have greater textual similarity (\textit{content preference}) and project description similarity (\textit{domain preference}) with their own historically resolved issues compared to FIs of other newcomers, measured using the median cosine and Jaccard similarity. 
These findings indicate that FIs are more similar to newcomers' historically resolved issues than to others' FIs. Thus, recommending issues to newcomers that are similar to their previous contributions may be more efficient and appropriate than suggesting identical issues to all newcomers.

\textbf{Newcomers' general OSS experience, activeness, and sentiment influence their FI selection.}
The results indicate that 19 features associated with the 37 newcomers' general OSS experience, activeness, and sentiment are related to the newcomers' FI selection. In this section, we use the number of commits, the number of commits within the past month, and the mean of the textual sentiment polarity values for reported issues as example features for the three feature groups to show their relationship with newcomers' FI choices. More results are included in the replication package.

Figure~\ref{typedis} reveals that newcomers who have made more commits and are more active recently than their peers tend to choose document rather than code issues.  
Figure~\ref{typedis} also shows that newcomers with less coding experience and more negative sentiment characteristics than their peers prefer adaptive rather than corrective or perfective issues. 
Moreover, Figure~\ref{domaindis} reveals that newcomers with more coding experience and greater activeness tend to resolve FIs of system software and software tool projects. Newcomers who contributed to FIs of web library and framework projects tend to exhibit more positive sentiment characteristics. 

\begin{result-rq}{Summary for RQ2:} 
Newcomers' expertise preference, and their general OSS experiences, activeness, and sentiment characteristics are related to their choices of FIs.
\end{result-rq}

\section{The PFIRec implementation and evaluation}\label{rq3}
Building on the insights of RQ1 and the identified influential factors of RQ2, we propose a personalized ranking model named Personalized First Issue Recommender (PFIRec), which aims to provide a specific newcomer in new projects with a tailored ranking of candidate issues that are most suitable for the newcomer. In this section, we formulate the problem, explain how the model is constructed, trained, and evaluated, and present the evaluation results. 

\subsection{Problem Formulation}
The primary goal of this research is to develop an automated recommendation system that suggests appropriate FIs in a new project for a given newcomer $d_i\in D=\{d_1,\ldots,d_I\}$ from a set of candidate issues $S=\{s_1,\ldots,s_N\}$ available in the project. Our aim is to rank each issue in order of the probability that it is suitable for the respective newcomer $d_i$. For each issue $s_n$, relevant factors can be extracted for the issue-newcomer pair $(s_n,d_i)$ to construct an input vector $\bm{x}_{n,i}$. We formulate this as an automated new project first issue recommendation problem: $y_{n,i}=f(\bm{x}_{n,i})\in [0,1]$, where $y_{n,i}$ represents the probability that the issue $s_n$ can be resolved by the newcomer $d_i$, and $\bm{x}_{n,i}\in \mathbb{R}^c$ denotes the input vector comprising $c$ features. These features include both characteristics influencing a newcomer's choice of their FI and characteristics of candidate issues.
The objective is to learn the model $f(\cdot)$ using training pairs $(\bm{x}_{n,i},y_{n,i})$ constructed from historical issue-newcomer pairs $(s_n,d_i)\in S\times D$, then utilize the trained model to predict the probabilities that candidate issues are suitable for a given newcomer. Finally, the candidate issues are recommended in descending order of probability for the newcomer.

\subsection{Model Training Data Specification}
To train the model $f(\cdot)$, we need to construct an input vector $\bm{x}_{n,i}$ for each issue-newcomer pair $(s_n,d_i)$ by pairing each newcomer with each candidate issue and extracting relevant factors.
Regarding $\bm{x}_{n,i}\in \mathbb{R}^c$, we follow the findings of RQ2 and collect all newcomer-related features presented in Section~\ref{datacollection-RQ2}. 

In addition, to build an effective model for issue recommendation, it is also important to extract relevant features from candidate issues. Previous studies~\cite{DBLP:conf/icse/XiaoHXTDZ22,DBLP:conf/sigsoft/HeSXHZ22}, have identified features that may indicate the difficulty, importance, and urgency of an issue based on its content and background. Therefore, we collect the following issue-related features:

\textbf{Content of candidate issues.} The content-related features consist of sentiment polarity, readability metrics (including Coleman-Liau formula, Kincaid Grade Level, Flesch Reading Ease, and Automated Readability Index), the length of the issue title, description, number of code snippets, URLs, images, the total number of labels, and the numbers of labels belonging to the 12 label categories according to~\cite{DBLP:conf/icse/XiaoHXTDZ22}: \Category{Bug, Documentation, Test, Build, Enhancement, Coding, New Feature, Newcomer-friendly} (e.g., \textit{good first issue}, \textit{easy} labels), \Category{Medium Difficulty, Difficult, Triaged, Untriaged}. These features of issues may indicate their type and difficulty.

\textbf{Background of candidate issues.} 
These features include: 
\begin{enumerate}
    \item Information of the project to which the issue belongs: number and ratio of open issues, number and ratio of GFIs, number of commits, PRs, closed issues, stars, and commit contributors, all of which can indicate whether the tasks provided by the project are sufficient for newcomers and whether the project is likely to attract many newcomers to compete for a candidate issue;
    \item Profile of the issue reporter: number of commits, PRs, reviewed PRs, reported issues, the ratio of GFIs among all reported issues in the project, maximum stars of contributed projects, and whether the reporter has commits in the project, all of which measures their experience and may be related to the difficulty of candidate issues~\cite{DBLP:conf/icse/XiaoHXTDZ22};
    \item Profile of the project owner, using the identical metrics as those for the issue reporter.
\end{enumerate}

\subsection{Ranking Model}
\label{ML-approach}
To rank the candidate issues based on the newcomers' individual features, we use LambdaMART~\cite{burges2010ranknet}, a state-of-the-art learning-to-rank model, as our ML model for training and testing.  
LambdaMART is a well-established machine learning technique that has demonstrated its effectiveness in software engineering sorting tasks~\cite{DBLP:journals/ase/CaoTLL18,DBLP:journals/tse/ZhouYCHMG22}. It is the MART (Multiple Additive Regression Tree)-based boosted tree~\cite{friedman2001greedy} version of LambdaRank~\cite{DBLP:conf/nips/BurgesRL06}, which is a learning-to-rank algorithm derived from RankNet~\cite{DBLP:conf/icml/BurgesSRLDHH05}. LambdaRank redefines the gradient of the objective function and considers ranking evaluators to improve the performance of RankNet. The use of MART in the LambdaMART algorithm enables it to perform gradient descent in the function space. The prediction value is a linear combination of the outputs of the regression trees~\cite{burges2010ranknet}.
Our experiment results demonstrate the superior effectiveness of LambdaMART in sorting issues for newcomers.

\subsection{Model Training}\label{sec:dataset} 
\begin{figure}
    \centering
    \includegraphics[scale=0.42,trim=0 0 0 0]{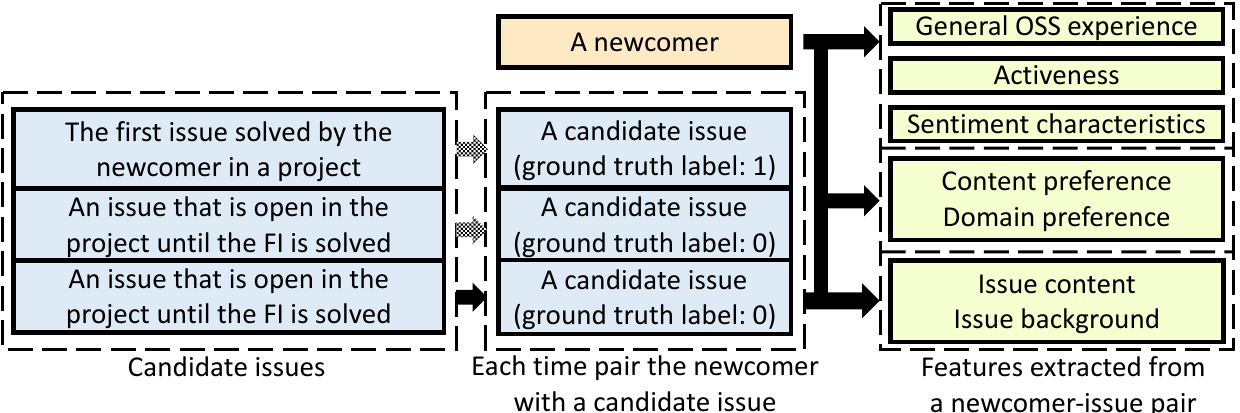}
    \caption{Feature extraction and ground truth setting.}
\vspace{-3mm}
    \label{fig:features}
\end{figure}

\textbf{Data collection and labeling}. To prepare the training data, we create lists that contain the features (inputs) and ground truth labels (outputs) of candidate issue-newcomer pairs. This is achieved by extracting all features ($\bm{x}_{n,i}$) and assigning the corresponding ground truth label $y_{n,i}$ for issue $s_n$ and newcomer $d_i$ via the process illustrated in Figure~\ref{fig:features}.

We consider FIs resolved in our chosen projects along with other issues that remain open until the FIs are resolved as candidate issues for the resolver of the FIs. Thus, for each of the 13,266 FIs identified in Section~\ref{datacollection-RQ1}, we constructed a candidate issue list. 
For each list, we pair each candidate issue with the resolver of the corresponding FI. To obtain the aforementioned features for each pair, we extract newcomer data using the GitHub GraphQL API~\cite{GitHubGr18:online} and issue data from the GFI-Bot dataset. The pair corresponding to the FI is labeled one, while the remaining pairs of candidate issues are labeled zero.

As the automatic issue recommendation approach has limited practical value for projects with only a few candidate issues, we exclude the lists from projects containing fewer than ten candidate issues. This exclusion accounts for approximately one-eighth of the total list count in our dataset, resulting in 11,615 lists covering a total of 68,858 issues (with some issues appearing in multiple lists). The median number of candidate issues per list is 32.

\begin{figure}
    \centering
    \includegraphics[scale=0.38,trim=0 0 0 0]{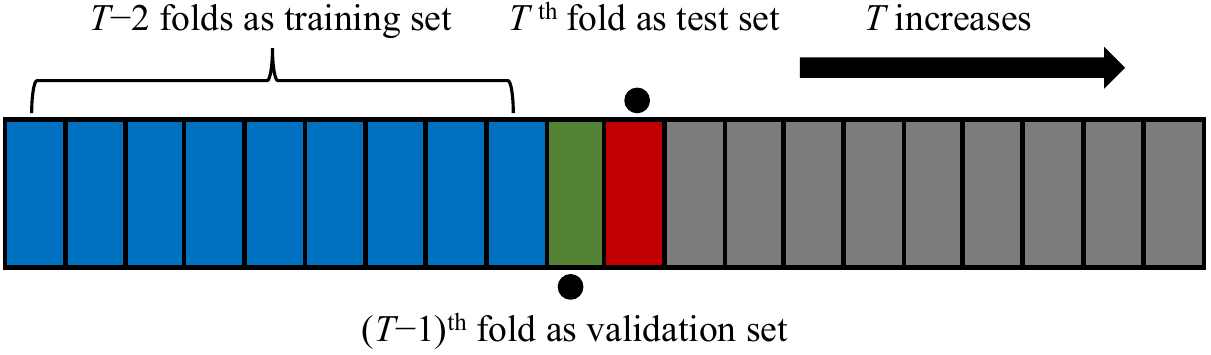}
    \caption{Building the datasets with the chronological FIs.}
\vspace{-3mm}
    \label{fig:sortt}
\end{figure}

\textbf{Process}. The practical application scenario envisioned for PFIRec involves utilizing training data collected based on historically resolved issues to train a model for ranking opening candidate issues. After some opening candidate issues are resolved regularly, PFIRec can update the training data to include newly resolved issues and retrain the model. To simulate such scenarios, we adopt a longitudinal data configuration that has been widely adopted in related studies~\cite{DBLP:conf/sigsoft/TamrawiNAN11,DBLP:conf/icse/WangY0HWW20,DBLP:conf/sigsoft/HeSXHZ22}. In our study, we sort the 11,615 lists based on their corresponding FIs' closing time in descending order and divide them into 20 equally sized folds (with additional lists included in the last fold).
As illustrated in Figure~\ref{fig:sortt}, we designate the first $T-2$ fold(s) as the training set, the $(T-1)$\textsuperscript{th} fold as the validation set, and the $T$\textsuperscript{th} fold as the test set. Then, we slide the window one fold and repeat the process for $T$ values ranging from 3 to 20 to train, validate, and test 18 models. For each model, we use its corresponding test set to evaluate the model's performance. 

Using the same feature extraction process as for the training and validation sets, we collect features for each newcomer in the test set and predict the likelihood of the newcomer resolving each candidate issue with the trained model. The resulting probabilities are used to rank the candidate issues in descending order. The ranked issues are presented by PFIRec and are evaluated.

\subsection{Evaluation}
In the evaluation, we aim to answer the following three evaluation questions (EQs). 
\begin{itemize}
    \item \textbf{EQ1:} How does the performance of PFIRec compare with its variants that employ different language and ranking models? This is to justify our choice of language and ranking model to build PFIRec.

    \item \textbf{EQ2:} How does the effectiveness of PFIRec in recommending issues to newcomers compare to the baselines? The purpose is to evaluate the performance improvement of PFIRec compared to state-of-the-art approaches.  

    \item \textbf{EQ3:} How sensitive is PFIRec to different feature groups and cross-project scenarios? The purpose is to gain a deeper understanding of the impact of various feature groups on prediction and evaluate PFIRec's generalizability in cross-project scenarios. 
\end{itemize}

\textbf{Evaluation Metrics}.
The top-$k$ recall ($R@k$) and FirstHit ($FH$) are popular ranking metrics to evaluate recommendation performance~\cite{DBLP:conf/wcre/SiowGFC020,DBLP:conf/recsys/YangSGL12,DBLP:conf/icse/WangY0HWW20}. 
We employ these metrics to measure how early the models can recommend the ``correct" FIs. 
Specifically, the top-$k$ recall calculates the ratio of successfully predicted ``correct" FIs to the total number of ``correct" FIs in the test set: $R@k=\frac{\sum_{i=1}^{m}n_i@k}{\sum_{i=1}^{m}n_i}$, where $n_i@k$ denotes the number of ``correct" FIs in the top-$k$ of the recommendation list for the $i$\textsuperscript{th} newcomer and $n_i$ represents the number of ``correct" FIs for the $i$\textsuperscript{th} newcomer. Regarding our problem, only one issue in a candidate issue list is the newcomer's ``correct" FI, so the top-$k$ recall simplifies to $R@k=1$ if the matching FI is ranked in the top $k$ of the recommendation list, and $R@k=0$ otherwise.
PFIRec aims to assist newcomers in identifying suitable issues at an early stage, thereby minimizing their erroneous attempts. We use $FH$ (FirstHit) as a metric to determine how early newcomers can identify suitable issues using PFIRec.
$FH$ records the median rank of the ``correct" FI in each recommendation list, and $FH-1$ represents the number of possibly unsuccessful attempts by the newcomer before finding a suitable issue.

\subsection{EQ1 Results}\label{evaluation-eq1}
To demonstrate the competitiveness of SimCSE and LambdaMART for our problem compared to other BERT-based language models and non-dedicated learning-to-rank models, we compare PFIRec's performance with various variants that use different language and ML models. 

\begin{figure}
    \centering
    \includegraphics[scale=0.42,trim=0 0 0 0]{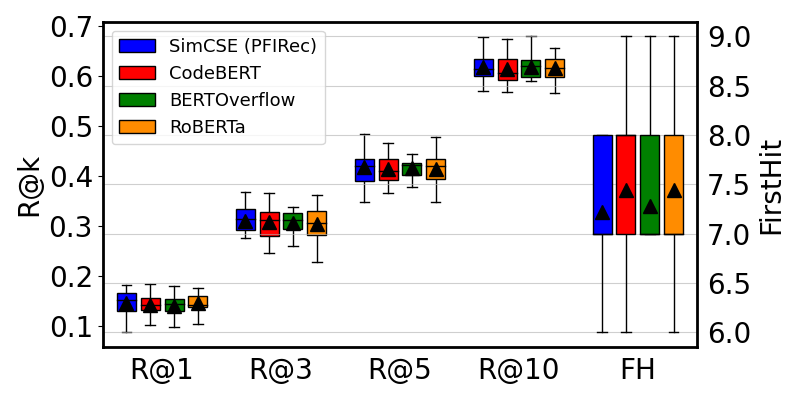}
    \caption{Performance of different language models.}
\vspace{-3mm}
    \label{berts}
\end{figure}

As introduced in Section~\ref{datacollection-RQ2}, we employ a BERT-based pre-trained language model for text embedding extraction. We compare PFIRec's ranking performance using SimCSE and the following three other BERT-based pre-trained models as feature extractors in Figure~\ref{berts}. 

\begin{itemize}[leftmargin=15pt]
\item \textbf{RoBERTa}~\cite{DBLP:journals/corr/abs-1907-11692} is one of the most popular BERT-based language models pre-trained with larger batch size and learning rates than the original BERT.
\item \textbf{CodeBERT}~\cite{DBLP:conf/emnlp/FengGTDFGS0LJZ20} is pre-trained based on the checkpoint of RoBERTa with the CodeSearchNet dataset~\cite{DBLP:journals/corr/abs-1909-09436} and can be used for both natural and programming languages. 
\item \textbf{BERTOverflow}~\cite{DBLP:conf/acl/TabassumMXR20} is pre-trained with sentences and tokens from Stack Overflow and has demonstrated effectiveness in the software engineering domain.
\end{itemize}

It can be seen that all four models perform similarly and that the SimCSE-based model achieves the best performance on $FH$, likely due to its optimization for similarity measurement problems.

\begin{figure}
    \centering
    \includegraphics[scale=0.42,trim=0 0 0 0]{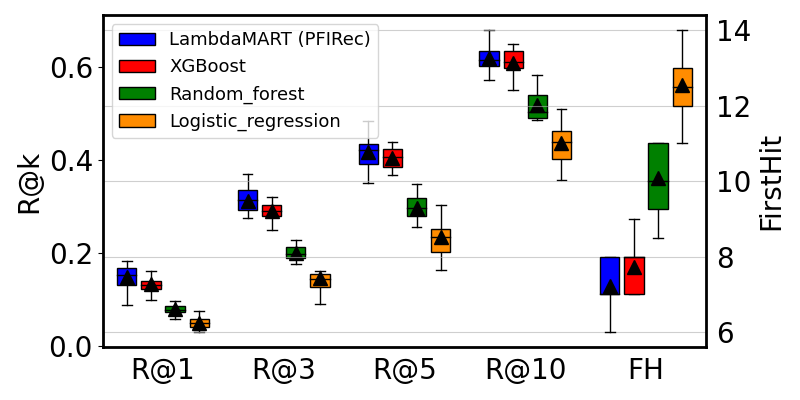}
    \caption{Performance of different ML models.}
\vspace{-3mm}
    \label{baseline0}
\end{figure}

\begin{figure}
    \centering
    \includegraphics[scale=0.42,trim=0 0 0 0]{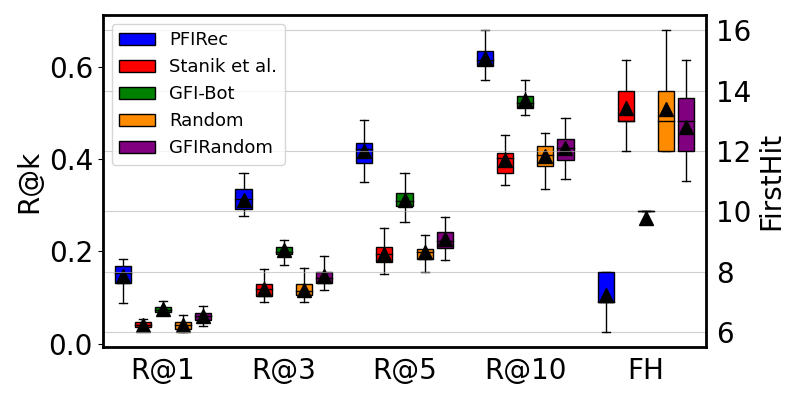}
    \caption{Performance of PFIRec and baselines.}
\vspace{-3mm}
    \label{baseline1}
\end{figure}

As explained in Section~\ref{ML-approach}, we employ LambdaMART as the machine learning model to rank candidate issues. An alternate method is to utilize a non-dedicated learning-to-rank model to estimate the probability of issues being suitable for specific newcomers and rank candidate issues based on the probability. 
In Figure~\ref{baseline0}, we compare LambdaMART's predictive performance with that of three other ML models, including the eXtreme Gradient Boosting (XGBoost) classifier~\cite{DBLP:conf/kdd/ChenG16}, random forest model, and logistic regression model from the Python scikit-learn package~\cite{DBLP:journals/jmlr/PedregosaVGMTGBPWDVPCBPD11}. The XGBoost model has the best performance among the three models, but is still slightly inferior to LambdaMART.
The median $FH$ of LambdaMART is 7, whereas for XGBoost it is 8. Thus, under our problem setting, the learn-to-rank model LambdaMART is the preferred choice over the other ML models.

\begin{result-rq}{Summary for EQ1:} 
PFIRec, which utilizes the SimCSE language model and the LambdaMART ML model, shows superior performance in ranking candidate issues compared to its variants employing other language models and ML models.
\end{result-rq}

\subsection{EQ2 Results}\label{baselines}
To evaluate the effectiveness of PFIRec in addressing the challenge of finding appropriate issues for specific newcomers, we compare it with the following four baselines.
\begin{itemize}[leftmargin=15pt]
\item
\textbf{Stanik et al.~\cite{DBLP:conf/icsm/StanikMMFM18}:} This approach leverages a random forest-based model that considers issue title and description length, sentiment score, and term frequency to identify suitable issues for newcomers.
\item
\textbf{GFI-Bot~\cite{DBLP:conf/sigsoft/HeSXHZ22}:} This is an under-development GitHub bot for automatically recommending GFIs based on RecGFI~\cite{DBLP:conf/icse/XiaoHXTDZ22}, which is a state-of-the-art automated approach. It predicts the probability that an issue suits newcomers and provides a ranked list of issues based on the probability. We do not choose RecGFI as a baseline due to its alignment with GFI-Bot's deployed model. 
\item
\textbf{Random:} This corresponds to randomly ranking candidate issues, which is equivalent to newcomers' practice of randomly checking open issues to identify appropriate ones.
\item
\textbf{GFIRandom:} This corresponds to the practice of randomly checking open issues with GFI labels first, and then randomly checking the remaining ones.
\end{itemize}

As we do not have access to the modified files and code to resolve the issues in our scenario, we do not compare PFIRec with the automatic GFI recommendation approach proposed by Huang et al.~\cite{DBLP:conf/esem/HuangWWLWW21}, which incorporates features of the modification information from the files and code.

Figure~\ref{baseline1} shows that the GFIRandom model only marginally outperforms the Random model, suggesting that manually-labeled GFIs may have limited effectiveness in assisting newcomers in issue selection. This aligns with the findings of Tan et al.~\cite{2020FSE-Tan-First}, which suggest that the problem of insufficient GFIs limits the GFI mechanism's effectiveness. Among the four baselines, GFI-Bot demonstrates the best performance. 
Compared to GFI-Bot, PFIRec reduces the median $FH$ by 3, indicating that it can \textbf{reduce one-third (6 vs. 9) of newcomers' unsuccessful attempts to find a FI in the median.}
Besides, compared with GFI-Bot, PFIRec improves the median $R@1,3,5,10$ by 114.3\% (0.15 vs. 0.07), 55.0\% (0.31 vs. 0.20), 35.5\% (0.42 vs. 0.31), and 17.3\% (0.61 vs. 0.52), respectively. Compared to the best practice without utilizing automated recommendation tools (i.e., GFIRandom), PFIRec can reduce the median of FH by 6, and increase $R@1,3,5,10$ by 150.0\% (0.15 vs. 0.06), 121.4\% (0.31 vs. 0.14), 90.9\% (0.42 vs. 0.22), and 45.2\% (0.61 vs. 0.42), respectively.

\begin{result-rq}{Summary for EQ2:} 
Compared to state-of-the-art approaches, PFIRec doubles the probability of the top-recommended issue being suitable for a specific newcomer, while reducing one-third of a newcomer's unsuccessful attempts to identify suitable first issues in the median.
\end{result-rq}

\subsection{EQ3 Results}\label{sensitivity}
To evaluate the impact of each feature group on the model's predictive performance, we construct seven variants of the model by removing the \textit{Content preference } (\textbf{noCont}), \textit{Domain preference} (\textbf{noDom}), \textit{General OSS experience} (\textbf{noGener}), \textit{Activeness} (\textbf{noAct}), \textit{Sentiment} (\textbf{noSenti}), \textit{Issue content} (\textbf{noIssCont}), and \textit{Issue background} (\textbf{noIssBack}) groups.  
Since the issues are collected from the 100 projects introduced in Section~\ref{methrq1}, we can conduct cross-project experiments to assess the model's generalizability to new projects, i.e., the prediction performance in projects outside the training set. To achieve this, we collect training and validation data from 90 projects and test data from the remaining ten projects. We perform a ten-fold cross-validation on the 100 projects and record its mean as the result. The other experimental steps are the same as described in Section~\ref{sec:dataset}.

Table~\ref{tab:abl} illustrates that removing any group leads to a retrograde alteration in the placement of the ``correct" FI in the recommendation list (resulting in an elevated $FH$) and a decrease in the values of $R@1,3,5,10$. 
Removing \textit{Issue content}, \textit{Issue background}, and \textit{General OSS experience} has the most significant impact, resulting in an increase in the median $FH$ by 4, 1, and 1, respectively.
When PFIRec is applied to the cross-project prediction scenario, a marginal shift is observed in the five evaluation metrics, indicating the generalizability of PFIRec to new projects.

\begin{table}[h!t]
\scriptsize
\centering
\vspace{-0.2cm}
\caption{Performance (mean/median) of variants of PFIRec and PFIRec for the cross-project scenario.}
\vspace{-0.2cm}
\label{tab:abl}
\begin{tabular}{p{1.2cm} p{1.1cm} p{1.1cm} p{1.1cm} p{1.1cm} p{0.8cm}}
\toprule
Models&R@1&R@3&R@5&R@10&FH\\
\midrule
noCont&0.14/0.14&0.30/0.29&0.41/0.41&0.61/0.61&7.6/7.5\\
noDom&0.14/0.15&0.31/0.31&0.42/0.42&0.61/0.62&7.3/7\\
noGener&0.13/0.12&0.28/0.28&0.40/0.39&0.60/0.60&7.7/8\\
noAct&0.14/0.14&0.31/0.31&0.41/0.41&0.61/0.60&7.4/7\\
noSenti&0.15/0.15&0.31/0.31&0.43/0.43&0.62/0.61&7.3/7.5\\
noIssCont&0.14/0.15&0.30/0.30&0.40/0.40&0.60/0.59&7.8/8\\
noIssBack&0.07/0.08&0.19/0.19&0.28/0.28&0.49/0.48&10.9/11\\
\midrule
CrossPro&0.15/0.16&0.32/0.32&0.44/0.44&0.63/0.63&7.5/7.5\\
\midrule
\textbf{PFIRec}&0.15/0.15&0.31/0.31&0.42/0.42&0.62/0.61&7.2/7\\
\bottomrule
\end{tabular}
\vspace{-0.3cm}
\end{table}

\begin{result-rq}{Summary for EQ3:} 
The removal of \textit{Issue content}, \textit{Issue background}, and \textit{General OSS experience} result in the most significant decrease in model performance. PFIRec maintains its generalizability in cross-project prediction scenario.
\end{result-rq}

\section{Discussion}\label{discussion}

\subsection{Comparison with Related Work}
Previous studies~\cite{DBLP:journals/tosem/AnvikM11,DBLP:conf/sigsoft/AshokJLRSV09,DBLP:conf/cikm/MacdonaldO06,DBLP:conf/icse/WangY0HWW20} on matching specific tasks with individual developers focus on recommending experts for tasks. 
On the contrary, our study aims to provide personalized task recommendations to individual project newcomers, who have not contributed code to the project, to assist in their onboarding process. To our knowledge, this is the first study to explore newcomers' personalized preferences in selecting the first issues to resolve in new projects. 
Adopting the PFIRec approach can provide two main benefits for projects. Firstly, it can help interested newcomers in finding suitable issues earlier and increase their chances of successfully joining the project. Secondly, this approach can reduce the likelihood of unsuitable newcomers attempting GFIs, allowing the issues to attract and retain matched newcomers. Reducing the number of attempts made by unsuitable newcomers can decrease the likelihood of rejected PRs and unfavorable communication, which in turn helps maintain their motivation to participate in the project. Furthermore, our findings emphasize the importance of experience matching in recommending tasks for newcomers. Issue reporters and project maintainers can consider practices such as providing task-type labels and listing required skills for GFIs to alleviate the difficulty faced by newcomers in finding suitable issues.

Our study can offer insight into future task-developer matching research: 1) Recommendation context modeling. 
Previous studies~\cite{DBLP:conf/kbse/Ye0WW18,DBLP:conf/icws/ZhouWS21} use only skill proficiency, collaboration history, and contributed projects' topics to model developers' contexts and do not use a learn-to-rank model. Our work confirms that general OSS experience accumulated by developers and the task's attribute features are useful for the recommendation model in addition to the previously used features; 2) Language model. Our results demonstrate the capability of the SimCSE language model based on BERT optimized for similarity metrics in computing text similarity; and 3) Ranking model. Although Wang et al.~\cite{DBLP:conf/icse/WangY0HWW20} also use LambdaMART to rank the recommendation list, our work shows that LambdaMART outperforms other non-learn-to-rank models in the recommendation scenario. 

Although our personalized recommendation objectives deviate from those for personalized recommendations of collaborators and repositories in software engineering~\cite{DBLP:conf/kbse/Ye0WW18,DBLP:conf/icws/ZhouWS21}, our evaluation of distinct language models and ranking models can provide valuable insights for future research endeavors in their research direction.

Compared to the state-of-the-art automatic GFI recommendation approach, searching for issues based on the recommendation lists provided by PFIRec can significantly reduce newcomers' unsuitable attempts by up to one-third in the median before finding a suitable issue. In fact, PFIRec doubles the success rate of newcomers who only check the first recommended issue.

\subsection{Threats to Validity}
\textbf{Internal Validity.} Firstly, some issues on GitHub may not be task-related~\cite{DBLP:conf/icsm/KallisSCP19}. PFIRec cannot distinguish whether an issue is task-associated or not. We suggest filtering out non-task issues by manual or automated methods to mitigate this threat. Filtering out non-task issues based on labels can be an effective strategy for projects that employ well-organized issue labeling. Secondly, following previous work on automatically recommending GFIs~\cite{DBLP:conf/icse/XiaoHXTDZ22}, we only consider the issues resolved by newcomers to be suitable for them. However, some issues may be suitable for particular newcomers but not seen by them or resolved by someone else first, resulting in incomplete positive labeling. However, we ensure that existing positive instances are correctly labeled. This approach is more appropriate than labeling data based solely on GFI tags, as it addresses problems regarding sparse GFI tags and potential mislabeling.
Thirdly, newcomers without GitHub experience may have different preferences for FIs, but PFIRec may recommend the same issues to them. Collecting newcomer experience data from sources beyond GitHub can mitigate this risk. Fourthly, the dataset may miss issues resolved by newcomers if they are not associated with commits or PRs. A tool that automatically associates commits, PRs, and the corresponding issues can be used to reduce this risk. Finally, our study treats developers who have not made any contribution to the project as newcomers. However, developers who have made several contributions may still want to search for suitable newcomer issues~\cite{2020FSE-Tan-First}. This can be addressed by adjusting the ground truth of the dataset according to the definition of newcomers (e.g., developers with less than a threshold number of commits in the project are considered newcomers).

\textbf{External Validity.} The results of RQ1 and RQ2 are based on a sample of 37 newcomers with multiple FIs, and the conclusions may not be generalized to other newcomers. 
The effectiveness of the identified features in predicting suitable FIs in a larger number of newcomers in Section~\ref{rq3} may partially mitigate this threat. Future work can investigate the selection of personalized first issues from more newcomers to verify our findings. Furthermore, our experiments are carried out on 100 GitHub projects, and the performance of PFIRec on other projects is unknown. To reduce this risk, we performed cross-project experiments, demonstrating the high generalization of PFIRec.

\section{Conclusion}\label{conclusion}
Newcomers to OSS projects often rely on GFI labels to identify issues that may suit them. Although some studies propose approaches to tag GFIs automatically, they do not consider newcomers' diverse backgrounds, which may result in a mismatch between their preferences and the identified issues. In this study, we analyzed the contributions of newcomers who have successfully resolved issues in multiple new projects and found that the issues resolved by them in new projects are more similar to their historical contributions than those resolved by other newcomers. This highlights the need for a personalized approach to issue recommendations to newcomers. By leveraging newcomer-related features from four feature groups (expertise preference, general OSS experience, activeness, and sentiment) along with variables encompassing the content and background of candidate issues, we propose PFIRec. This novel approach uses a LambdaMART ranking model to deliver personalized issue recommendations specifically tailored for newcomers. The effectiveness of PFIRec is experimentally evaluated and compared with existing approaches that recommend issues to newcomers, demonstrating its superior performance. 

Future research can explore the trajectories of newcomers on GitHub and their resolved first issues, to further uncover patterns in their first issues selection. It is also essential to investigate how diverse patterns of first issue selection impact the subsequent participation of newcomers from different backgrounds in projects. These studies have the potential to improve personalized first issue recommenders and provide best practice guidelines for newcomers when choosing first issues.

\section*{Acknowledgment}
This work is supported by the National Natural Science Foundation of China Grant 61825201 and 62332001.

\bibliographystyle{IEEEtran}
\bibliography{references}

\end{document}